\documentclass[sigconf]{acmart}
\AtBeginDocument{%
  }

\setcopyright{acmlicensed}
\copyrightyear{2018}
\acmYear{2018}
\acmDOI{XXXXXXX.XXXXXXX}
\acmConference[ACM CCS'25]{ACM Conference on Computer and Communications Security (CCS)}{October 13--17, 2025}{Taipei, Taiwan}
\acmISBN{978-1-4503-XXXX-X/2018/06}

\usepackage{enumitem}
\usepackage{soul}          
\usepackage[most]{tcolorbox}
\usepackage{booktabs}
\usepackage{pgfplots}
\pgfplotsset{compat=1.18}  
\usepackage{pdfpages}      

\pgfplotsset{compat=1.17}
\pgfplotsset{compat=1.18}

\usepackage{tikz}
\usetikzlibrary{shapes.geometric, arrows.meta, positioning,calc,fit}

\tikzstyle{uxbox} = [rectangle, rounded corners, minimum width=3.2cm, minimum height=1.4cm, text centered, draw=gray!80, fill=gray!10, font=\small]

\tikzstyle{arrow} = [thick,->,>=Stealth]

\tikzstyle{module} = [rectangle, rounded corners=3pt, minimum width=3.5cm, minimum height=1.2cm, text centered, draw=black, line width=0.8pt, fill=gray!10]

\tikzstyle{output} = [rectangle, draw=black, rounded corners=3pt, minimum width=3.5cm, minimum height=1.2cm, text centered, fill=white]

\begin{document}

\title{\textit{Too Much to Trust?} Measuring the Security and Cognitive Impacts of Explainability in AI-Driven SOCs}

\author{Nidhi Rastogi}
\authornote{nidhi.rastogi@rit.edu, corresponding author}
\orcid{1234-5678-9012}
\affiliation{%
  \institution{Rochester Institute of Technology}
  \city{Rochester}
  \state{New York}
  \country{USA}
}

\author{Shirid Pant}
\affiliation{%
  \institution{Rochester Institute of Technology}
  \city{Rochester}
  \state{New York}
  \country{USA}
}

\author{Devang Dhanuka}
\affiliation{%
  \institution{Rochester Institute of Technology}
  \city{Rochester}
  \state{New York}
  \country{USA}
}

\author{Amulya Saxena}
\affiliation{%
  \institution{Rochester Institute of Technology}
  \city{Rochester}
  \state{New York}
  \country{USA}
}

\author{Pranjal Mairal}
\affiliation{%
  \institution{Independent Researcher}
  \city{Ahmedabad}
  \state{Gujarat}
  \country{India}
}

\renewcommand{\shortauthors}{Anonymous et al.}

\begin{abstract}
Explainable AI (XAI) holds significant promise for enhancing the transparency and trustworthiness of AI-driven threat detection in Security Operations Centers (SOCs). However, identifying the appropriate level and format of explanation, particularly in environments that demand rapid decision-making under high-stakes conditions, remains a complex and underexplored challenge. To address this gap, we conducted a three-month mixed-methods study combining an online survey (N1=248) with in-depth interviews (N2=24) to examine (1) how SOC analysts conceptualize AI-generated explanations and (2) which types of explanations are perceived as actionable and trustworthy across different analyst roles. Our findings reveal that participants were consistently willing to accept XAI outputs, even in cases of lower predictive accuracy, when explanations were perceived as relevant and evidence-backed. Analysts repeatedly emphasized the importance of understanding the rationale behind AI decisions, expressing a strong preference for contextual depth over a mere presentation of outcomes on dashboards. Building on these insights, this study re-evaluates current explanation methods within security contexts and demonstrates that role-aware, context-rich XAI designs aligned with SOC workflows can substantially improve practical utility. Such tailored explainability enhances analyst comprehension, increases triage efficiency, and supports more confident responses to evolving threats.
\end{abstract}




\maketitle

\section{Introduction}
\label{sec:intro}

Modern Security Operations Centers (SOCs) ingest a continuous stream of security alerts and notifications from monitoring tools (e.g., intrusion detection systems, SIEM platforms, Endpoint Detection and Response agents, etc.) that signal potential malicious activity\cite{hassan2019nodoze,gonzalez2021security}. Examples of security alerts include indications of a brute-force login attempt (e.g., repeated failed login attempts from an external IP address), detection of malware activity (e.g., malicious file execution or unusual outbound network connections), or identification of phishing attempts (e.g., suspicious email patterns and known malicious URLs). Such anomalous activities within the network may indicate a security threat, and these alerts require immediate analyst assessment to ascertain their validity, potential impact, and necessary actions.  
\par
However, while alert detection has become increasingly automated, interpreting and responding to these alerts remains a deeply human, cognitively demanding task~\cite{mink2023everybody}. SOC analysts should quickly determine the relevance and severity of alerts under intense time pressure, frequently without clear insights into the underlying reasoning behind the automated detection. This disconnect leads to \emph{alert fatigue}, a well-documented phenomenon marked by cognitive overload, overlooked threats, and inefficient triage workflows. Recent evidence shows that SOCs receive an average of 3,832 alerts per day, 62\% of which are ignored; over 70\% of analysts report feeling overwhelmed~\cite{hassan2019nodoze,gonzalez2021security,rastogi2022contextual}. Effectively managing alerts is, therefore, crucial since missing a real threat alert notification can be detrimental to an organization, whereas chasing too many false alerts can lead to wasting resource~\cite{yang2024true,alahmadi202299}.
\par
This points to a fundamental challenge between interpretability and usability in high-stakes, time-sensitive domains like security. On the one hand, explainable AI (XAI) techniques promise greater transparency, analyst trust, and insight into the reasoning behind AI-generated alerts~\cite{bhusal2023sok,rastogi2025survey,nadeem2023sok}. On the other hand, SOC analysts operate in a high-stakes, time-sensitive environment that demands any explanation be fast, context-relevant, and cognitively lightweight to avoid impeding response. We posit that an AI-provided explanation's usefulness highly depends on \emph{who} uses it, \emph{when} it is used, and \emph{why} someone consults it. An entry-level Tier-1 triage analyst, for example, might benefit from a concise summary of why an alert is likely malicious, whereas a Tier-3 threat hunter may require deeper technical details and model rationale~\cite{kersten2024security,yang2024true,maxam2024interview,alahmadi202299}. If explanations are too generic or too complex for the situation, they risk confusing the analyst or slowing down the investigation. Unfortunately, existing XAI methods, largely designed for technical debugging or regulatory transparency, fail to account for these contextual factors. As a result, explanations positioned in current research can either overwhelm analysts with extraneous details or, conversely, oversimplify important information, limiting their practical utility in the SOC workflow.
\par
While recent studies have explored Security Operations Centers (SOCs) and their integration with AI/ML tools, none have specifically examined how explainability of these models influences analyst workflows. This paper addresses this crucial and timely gap by providing the first comprehensive empirical investigation into the cognitive impacts and trust implications of AI-driven explanations within SOC environments.

\begin{figure}[ht]
    \centering
    \setlength{\fboxsep}{2pt} 
    \setlength{\fboxrule}{0.5pt} 
    \fbox{%
        \resizebox{.85\linewidth}{!}{%
            \includegraphics[width=.99\linewidth]{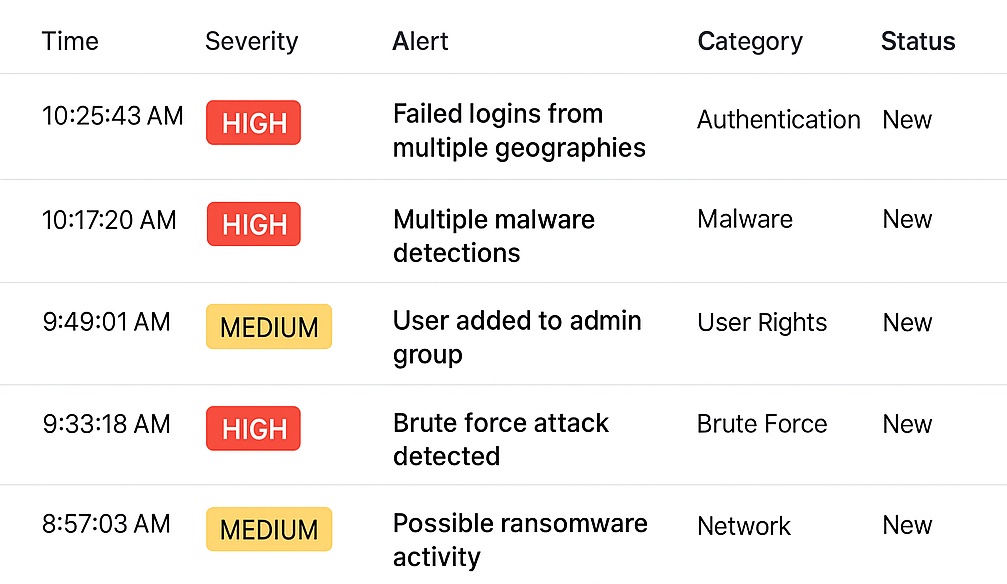}
        }
    }
    \caption{Illustrative SOC Dashboard with alerts of different categories - High and Medium.}
    \label{fig:soc-dashboard-alert}
\end{figure}

\begin{table}[ht]
\centering
\footnotesize
\begin{tabular}{c l l}
\toprule
\textbf{Tier} & \textbf{Role} & \textbf{Responsibilities} \\
\midrule
1 & Triage & Alert validation, severity assessment, escalation \\[2pt]
2 & Incident Response & Investigation, event correlation, attribution \\[2pt]
3 & Threat Hunting & Proactive hunting, forensics, rule refinement \\[2pt]
\bottomrule
\end{tabular}
\caption{SOC Analyst Tiers and Responsibilities}
\label{tab:soc_tiers_compact}
\end{table}

To address this, we introduce \textsf{CORTEX}- the \textit{\ul{CO}ntextual \ul{R}ole and \ul{T}rust aware \ul{EX}planation} framework for the AI model's explanations used in SOC. \textsf{CORTEX} tailors the model’s output as well as three key contextual dimensions: the analyst’s investigative tier (role), the operational context of the alert, and the current cognitive load of the task. The goal of CORTEX is to deliver the optimal information to the analyst at the right time. We formalize this intuition through the \textit{Trust-Explainability curve}, which models how explanation complexity and timing affect perceived utility and trust across different SOC roles. To evaluate and refine these ideas, we conducted a mixed-methods study over three months. We surveyed (N1=248) SOC professionals, followed by 24 in-depth interviews with analysts (over 810 minutes) across four continents, representing diverse roles, sectors (including managed security service providers, banking, IT, and finance), and years of experience (ranging from 1–7 years of on-the-job experience). The empirical study allowed us to examine the real-world limitations of established XAI methods (e.g., SHAP, LIME, LEMNA~\cite{ribeiro2016should,lundberg2017unified,guo2018lemna}) within realistic SOC contexts.

The findings revealed several challenges where generic explanations fall short, informing the design of our context-aware solution, \textsf{CORTEX}. Based on these insights, we propose a set of design recommendations for building more effective explainable security methods grounded in both the qualitative feedback from analysts and quantitative data on triage performance.
\par
Our study is guided by the following research questions:
\begin{itemize}[leftmargin=1.5em]
    \item \textbf{RQ1:} How do SOC analysts conceptualize explanations for AI-driven alerts, and which explanations are perceived as actionable or trustworthy across investigative roles?
    \item \textbf{RQ2:} Under which operational scenarios (e.g., active incident responses, routine monitoring, escalations to leadership) do analysts find explainability beneficial or neutral with respect to decision-making?
    \item \textbf{RQ3:} What technical, organizational, and cognitive barriers impede effective adoption and trust of XAI within SOC workflows?
\end{itemize}

\par
This research study makes the following key contributions:
\begin{enumerate}[leftmargin=1.5em]
    \item \textbf{We provide actionable design guidelines grounded in CORTEX,} enabling alerting systems to dynamically adjust explanations based on investigative tier, role, and operational urgency.
    \item \textbf{We introduce the \textit{Trust-Explainability curve},} a novel approach characterizing when and how explainability shifts from being helpful to burdensome in different investigative contexts.
    \item \textbf{We present a comprehensive empirical investigation,} including a 248-participant survey and 24 analyst interviews totaling 761 minutes across SOCs in North America, Europe, Africa, and Asia. Full instruments are provided here\footnote{\url{https://anonymous.4open.science/r/CCS2025-XAISocUsability-4B60/README.md}}.
\end{enumerate}



\section{Background and Related Work}\label{sec:background}

\begin{table*}[ht]
\centering
\small
\begin{tabular}{p{3cm}p{14cm}}
\toprule
\textbf{Explainability Model} & \textbf{Explanation of Network Log Example} \\
\midrule
\multicolumn{2}{l}{\textbf{Network Log Example:}} \\
\multicolumn{2}{l}{\texttt{Jan 15 10:41:17 host01 sshd[5021]: Failed password for invalid user 'admin' from IP 192.168.3.45 port 49538 ssh2}} \\
\midrule
LIME \cite{ribeiro2016should} & Highlights keywords or features such as \colorbox{gray!15}{\texttt{"Failed password''}}, \colorbox{gray!15}{\texttt{``invalid user''}}, and \colorbox{gray!15}{\texttt{``admin''}} as key contributors using simplified decision rules based on perturbed examples. \\[4pt]

SHAP \cite{lundberg2017unified} & Quantifies feature contributions numerically, assigning high Shapley values to features like \colorbox{gray!15}{\texttt{``invalid user''}}, \colorbox{gray!15}{\texttt{``admin''}}, and IP address \colorbox{gray!15}{\texttt{``192.168.3.45''}}, indicating their influence on the alert. \\[4pt]

LEMNA \cite{guo2018lemna} & Explains the detection using a local nonlinear approximation that highlights unusual credentials (\colorbox{gray!15}{\texttt{'admin'}}) and login failures from the IP (\colorbox{gray!15}{\texttt{192.168.3.45}}). \\[4pt]

XNIDS \cite{xu2023xnids} & Uses feature importance derived from neural networks specialized for network events, emphasizing repeated access attempts, unusual usernames (\colorbox{gray!15}{\texttt{'admin'}}), and rare IP addresses (\colorbox{gray!15}{\texttt{192.168.3.45}}). \\[4pt]

Grad-CAM \cite{selvaraju2017grad} & Visually emphasizes important segments of the log (e.g., \colorbox{gray!15}{\texttt{'admin'}} user and IP \colorbox{gray!15}{\texttt{192.168.3.45}}) through activation maps produced by neural networks. \\[4pt]

Integrated Gradients \cite{sundararajan2017axiomatic} & Traces incremental contributions of log elements (username \colorbox{gray!15}{\texttt{'admin'}}, phrase \colorbox{gray!15}{\texttt{``Failed password''}}, IP \colorbox{gray!15}{\texttt{192.168.3.45}}) directly attributing them to the alert decision. \\[4pt]

DeepLIFT \cite{shrikumar2017learning} & Highlights significant differences from normal activity such as \colorbox{gray!15}{\texttt{``invalid user''}}, username \colorbox{gray!15}{\texttt{'admin'}}, and suspicious IP address \colorbox{gray!15}{\texttt{192.168.3.45}}, identifying their strong relative contribution. \\[4pt]

Layer-wise Relevance Propagation (LRP) \cite{bach2015pixel} & Assigns relevance scores through network layers back to inputs like \colorbox{gray!15}{\texttt{``invalid user''}}, username \colorbox{gray!15}{\texttt{``admin''}}, and IP \colorbox{gray!15}{\texttt{``192.168.3.45''}}, clarifying their direct contribution to the alert. \\[4pt]
\bottomrule
\end{tabular}
\caption{Various Explanations generated by widely researched Explainability Models using a Security Network Log Example}
\label{tab:XAI-summary}
\end{table*}

\subsection{Explainability Methods}
\label{subsec:xai_methods}
Recent explainable AI (XAI) techniques can be broadly categorized into two classes: \textit{model-agnostic} and \textit{model-specific}, each with distinct implications for interpretability, fidelity, and operational utility~\cite{guidotti2018survey,doshi2017towards}. \textbf{Model-agnostic methods} operate independently of the underlying learning algorithm and are usually applied post-hoc to generate explanations for any ``black-box model.'' Some of the widely used models include LIME~\cite{ribeiro2016should} and SHAP~\cite{lundberg2017unified}, which approximate local model behavior by input perturbation and assign importance scores to features via surrogate models. These methods are valued for their flexibility and generality and often produce simplified, human-readable outputs such as ranked feature lists or visual heatmaps. 
Examples of the explanations generated by the above-mentioned models is provided in Table~\ref{tab:XAI-summary}. 
\textbf{Model-specific methods}, by contrast, are tightly coupled with the internal structure of a particular model to generate explanations. They exploit architectural details (weights, gradients, or activations) to produce explanations grounded in the model’s internal computation~\cite{samek2021explaining}. For example, Grad-CAM~\cite{selvaraju2017grad}, Integrated Gradients~\cite{sundararajan2017axiomatic}, DeepLIFT~\cite{shrikumar2017learning}, and Layer-wise Relevance Propagation~\cite{bach2015pixel} are commonly used in deep learning models, particularly for tasks in computer vision and medical diagnostics~\cite{caruana2015intelligible}. 
Beyond these, a few methods have been tailored for security analytics, such as LEMNA for intrusion detection explanations~\cite{guo2018lemna} and XNIDS~\cite{xu2023xnids} for network anomaly interpretation. While these security-specific XAI methods align explanations with security telemetry semantics, they remain insufficiently responsive to real-time SOC demands and analyst heterogeneity. Additionally, they assume a one-size-fits-all explanation format, overlooking context-specific needs across different operational scenarios and roles.

\subsection{Explainability and Human-centered Research in Security}
\label{subsec:explainability_in_security}
Existing literature underscores the theoretical promise of XAI for security~\cite{baruwal2024humanAI}, yet real-world deployments in SOCs remain rare and methodologically underexamined. Many studies prioritize model-centric evaluation metrics like accuracy or robustness without accounting for how explanations are actually perceived, trusted, or utilized by SOC analysts~\cite{apruzzese2018effectiveness, sharma2020mimic}. This leaves a critical blind spot: \textbf{an explanation that looks good in theory may be impractical or even counterproductive if it does not fit analysts’ cognitive workflows}. Empirical studies suggest that analysts seek features like uncertainty quantification, attribution trails, and causal narratives~\cite{rastogi2025survey}. However, these desiderata are infrequently incorporated into real-world systems, reflecting a persistent disconnect between academic prototypes and deployment-ready solutions. Furthermore, the cognitive demands placed on analysts are seldom considered when designing explanation objectives or formats, leading to explanations that are either too abstract or too burdensome to be actionable. This gap motivates a shift toward human-centered explainability approaches in security.
\par
Recent research on SOC AI/ML tools reveals areas for improvement in alert handling and analyst workflows. Kersten et al.~\cite{kersten2024security} introduced an Alert Investigation Support System for Tier-1 analysts, Mink et al.~\cite{mink2023everybody} found ML systems often lack context-rich, actionable explanations, and Yang et al.~\cite{yang2024true} highlighted inefficiency due to excessive irrelevant alerts. Maxam and Davis~\cite{maxam2024interview} underscored flexible, exploratory threat-hunting over rigid approaches, and Alahmadi et al.~\cite{alahmadi202299} emphasized reliable, explainable, and contextually valuable alerts. Yet none explicitly examines the explainability of AI/ML models in SOCs. Our study fills this gap by empirically investigating how model explanations impact analysts’ workload, trust, and decision-making. Given SOCs’ reliance on AI-driven detection, poor explanations risk undermining trust, efficiency, and security, making context-aware explanation frameworks imperative.

SOCs are fast-paced and cognitively demanding environments~\cite{kersten2024security,yang2024true,alahmadi202299}. Analysts should continuously synthesize incomplete threat intelligence, evolving attack indicators, and organizational protocols under significant time pressure. In this context, the utility of an explanation depends not only on its correctness but also on its \emph{relevance, brevity, and timing} with respect to the analyst’s task~\cite{bhusal2023sok, suh2024more}. From a cognitive load perspective, an explanation that is too verbose or poorly timed imposes an extraneous cognitive load on the analyst’s limited working memory~\cite{rastogi2025survey}. Traditional explanation mechanisms, such as saliency maps or ranked feature contributions, often fail to meet these criteria. Prior research in human-computer interaction highlights that role-sensitive, context-aware design significantly improves decision quality under pressure~\cite{caruana2015intelligible,carroll2003making}. 
\section{CORTEX}\label{sec: cortex}
\paragraph{Motivation}
To address these human factors, we conceptualize \textsf{CORTEX}- \textit{\ul{CO}ntextual \ul{R}ole and \ul{T}rust aware \ul{EX}planation} framework that dynamically adjusts the granularity and format of explanations to align with an analyst’s cognitive bandwidth, the incident severity, and their investigative role. 
In practice, this means providing concise, high-level justifications during time-constrained, high-pressure triage moments and richer, detailed explanations during in-depth investigations. By tailoring explainability to context, \textsf{CORTEX} aims to support appropriate trust in the AI system, informing the analyst just enough to calibrate trust without overloading them. This adaptive strategy is grounded in theories of cognitive load, which is presenting information in digestible chunks to avoid overwhelming the user. It also aligns with the idea of situated cognition that knowledge and explanations are most meaningful when presented in the context in which they are used.

\subsection{Trust-Explainability Curve}\label{subsec: trust_explainability}
\noindent
\paragraph{Trust-Explainability curve}
Most prior studies have overlooked the realities of SOC workflows and the diversity of analyst roles~\cite{rastogi2025survey}. Explanations are rarely evaluated across the full spectrum of analyst tiers (see Table~\ref{tab:soc_tiers_compact}) or integrated into actual SOC processes. This lack of deployment challenges limits effectiveness and can even erode trust. A context-agnostic explanation strategy might result in \emph{miscalibrated trust}: Tier-1 analysts could place blind faith in an AI’s assessment if given an authoritative-looking explanation they don’t fully understand (potential \emph{misuse} of automation), whereas Tier-3 experts might ignore using the AI’s advice if the explanations seem too simplistic or irrelevant to their needs~\cite{norman1990problem}. To address this gap, we introduce a structured, empirically grounded model of explanation utility, the \textbf{Trust–Explainability curve}, that explicitly captures how explanation complexity and timing affect user trust and decision efficacy across different SOC roles. This curve aims to align user trust with system reliability while considering how the content or complexity of explanations influences that trust. It posits that for each type of SOC analyst, there is an optimal level of explanation detail that maximizes the analyst’s calibrated trust in the AI; \textit{too little explanation can breed skepticism or misunderstanding, while too much can cause confusion and delay}.
\paragraph{Explanation friction} To describe the mismatch between an explanation’s form and the analyst’s immediate cognitive context, we introduce \textbf{explanation friction}. It arises when an explanation modality fails to scale with the situation’s urgency or the user’s expertise level. For example, a verbose textual explanation that might aid a Tier-3 forensic analyst can become an impediment to a Tier-1 triage analyst who needs to make a split-second decision. It can be measured via indicators such as prolonged alert handling times, excessive back-and-forth queries for clarification, or analyst confusion in user studies.

\subsection{CORTEX Framework}\label{sec:cortex_framework}


\begin{figure}[ht]
\centering
\resizebox{1.05\linewidth}{!}{%
\begin{tikzpicture}[%
  font=\small,
  node distance=10mm and 8mm,
  box/.style={draw, rounded corners, align=center, minimum width=2cm, minimum height=1.2cm},
  arrow/.style={->,>=stealth,thick}
]

\node[box, fill=gray!10] (profiler) {\textbf{User Profiler}\\ \& Access Controller};

\node[box, fill=gray!10, below=of profiler] (context) {\textbf{Context Engine}\\ \footnotesize (Threat Intel, Historical Data)};

\node[box, fill=gray!10, right=of profiler, xshift=6mm] (generator) {\textbf{Explanation Generator}\\ \footnotesize (Layered Detail)};

\node[box, fill=gray!10, below=of generator] (feedback) {\textbf{Feedback \& Learning}\\ \footnotesize (Refine Explanations)};

\draw[arrow] (profiler) -- (context);
\draw[arrow] (profiler) -- (generator);
\draw[arrow] (context) -- (generator);
\draw[arrow] (generator) -- (feedback);

\node[box, fill=gray!5, above=of profiler, yshift=-6mm] (alert) {\textit{Incoming Alert}\\ \textit{(Severity, Type, etc.)}};
\draw[arrow] (alert) -- (profiler);

\node[box, fill=gray!5, right=of feedback, xshift=7mm] (explanation) {\textit{Adaptive Explanation}\\ \textit{(Tier-based, Context-aware)}};
\draw[arrow] (feedback) -- (explanation);

\node[above left=2mm of profiler, align=left, font=\footnotesize] (label1) {\textbf{(1)}};
\node[above right=2mm of generator, align=left, font=\footnotesize] (label2) {\textbf{(3)}};
\node[below left=2mm of context, align=left, font=\footnotesize] (label3) {\textbf{(2)}};
\node[below right=2mm of feedback, align=left, font=\footnotesize] (label4) {\textbf{(4)}};

\end{tikzpicture}
}
\caption{Architecture of the \textsf{CORTEX} framework, integrating user tier, context, feedback, and provenance into adaptive explanation workflows.}
\label{fig:cortex_arch}
\end{figure}


\subsubsection{Core Principles}

\paragraph{1. Role-Sensitivity.} Building upon the tiered SOC model, \textsf{CORTEX} accommodates different investigative roles (e.g., Tier-1 vs. Tier-3) by delivering only the level of technical detail relevant to that role. For instance, a Tier-1 analyst sees high-level incident summaries and straightforward action recommendations, while a Tier-3 threat hunter can drill down into model weights, deep technical logs, or advanced correlation graphs. 

\paragraph{2. Contextual Adaptation.} \textsf{CORTEX} dynamically adjusts its explanatory detail based on: \begin{itemize}[leftmargin=1.15em] \item \emph{Alert Severity:} High-urgency incidents (e.g., active data exfiltration) trigger concise, actionable explanations, while lower-severity scenarios allow richer narratives and optional forensics. \item \emph{Operational Phase:} During rapid triage, the framework offers minimal but critical cues. Later, in forensic or root-cause analyses, deeper layers (e.g., historical correlation, model provenance) become readily accessible. \item \emph{Analyst Feedback:} The system tracks user interactions (e.g., which panels are expanded and which details are skipped) to learn personal preferences. Repeated behaviors guide the interface to emphasize or omit certain explanation layers in future alerts. \end{itemize}

By leveraging real-time cues, \textsf{CORTEX} addresses \emph{explanation friction} directly, avoiding static, one-size-fits-all descriptions that might overwhelm some analysts or under-inform others.

\paragraph{3. Trust-Aware Explanation Depth.} Reflecting insights from our Trust-Explainability curve, the framework carefully meters how much detail is revealed at once. 
\textsf{CORTEX} thus modulates transparency \emph{progressively}: essential indicators (e.g., suspicious file hashes, blacklisted IP addresses) appear immediately, while deeper rationales, such as Shapley value plots~\cite{lundberg2017unified} require an explicit role that typically requires advanced context.

\subsubsection{Framework Architecture}

Figure~\ref{fig:cortex_arch} illustrates \textsf{CORTEX}’s conceptual design, consisting of four primary modules:

\begin{enumerate}[leftmargin=1.25em] \item \textbf{User Profiler and Access Controller:} Maintains information on each analyst’s tier, privileges, and usage patterns. This module gates the level of detail provided, ensuring that explanations align with role-based permissions (e.g., Tier-3 can see rule-engine logs that Tier-1 cannot). \item \textbf{Context Engine:} Integrates external threat intelligence, historical incident data, and organizational knowledge to inform relevant references in the explanations. By grounding each alert in real-time context, \textsf{CORTEX} enhances perceived credibility and fosters trust. \item \textbf{Explanation Generator:} Dynamically assembles explanations with graduated complexity, ranging from short text summaries to deeper causal chains. It factors in current alert severity, user preference, and the \emph{explanation friction} thresholds to avoid overwhelming the analyst. \item \textbf{Feedback and Learning Module:} Observes how analysts interact with each explanation layer (e.g., which sub-panels are opened, how often advanced details are viewed). Over time, it refines default explanation levels on a per-user or per-role basis. \end{enumerate}

\subsubsection{Example Workflow} To illustrate how \textsf{CORTEX} operates in practice, consider a Tier-1 analyst investigating a suspicious outbound connection. The system initially presents: \begin{itemize}[leftmargin=1.25em] \item A concise alert summary: ``Outbound traffic to known malicious IP \texttt{198.51.100.23} flagged as suspicious.'' \item Immediate recommended steps: ``Quarantine host \texttt{host01}; escalate for deeper investigation.'' \end{itemize} Should the analyst escalate to Tier-2 or Tier-3, the interface seamlessly expands, revealing a timeline linking the alert to prior failed login attempts, any correlated intelligence from MITRE ATT\&CK patterns~\cite{strom2018mitre}, and optional model-specific details (e.g., local feature attributions). This layered explanation strategy directly mitigates the risk of \textbf{explanation friction}, enabling time-pressed Tier-1 staff to act swiftly while affording advanced analysts a richer investigative canvas.

\subsubsection{Addressing Explanation Friction and Trust Calibration} As detailed in Section~\ref{subsec: trust_explainability}, \emph{explanation friction} can impede effective SOC operations if not carefully managed. \textsf{CORTEX} counters this by providing graduated forms of transparency, ensuring no single user is inundated with irrelevant details. Guided by the Trust-Explainability curve, the system aims to deliver just enough rationale to maintain appropriate trust, where an analyst neither underestimates the system’s reliability nor overestimates its certainty.

\begin{table*}[ht]
\centering
\renewcommand{\arraystretch}{1.12}   
\begin{tabular}{p{0.5cm} p{6cm} p{9cm}}
\toprule
\textbf{Tier} &
\textbf{Observed Log Patterns / Features} &
\textbf{Key Analyst Actions \& Explanation Requirements} \\
\midrule
\textbf{1} & 
\begin{minipage}[t]{\linewidth}\raggedright
\begin{itemize}[leftmargin=*]
  \item \colorbox{gray!20}{Repeated VPN login failures} quickly followed by a
        \colorbox{gray!20}{successful login} from
        \colorbox{gray!20}{suspicious external IP (198.51.100.23)}.
  \item \colorbox{gray!20}{Alert for large data transfer} from
        \texttt{198.51.100.23} to \texttt{10.0.0.72}.
\end{itemize}
\end{minipage}
&
\begin{minipage}[t]{\linewidth}\raggedright
\begin{itemize}[leftmargin=*]
  \item \textbf{Rapid Triage}: validate or dismiss high‑volume routine alerts.
  \item \textbf{Severity Assessment}: decide if the pattern is malicious or benign.
  \item \textbf{Escalation}: suspicious external IP $\Rightarrow$ Tier‑2.
  \item \textbf{Explain}: concise, high‑level indicators for split‑second action.
\end{itemize}
\end{minipage}
\\
\midrule
\textbf{2} & 
\begin{minipage}[t]{\linewidth}\raggedright
\begin{itemize}[leftmargin=*]
  \item \textbf{Rapid Triage}: Validate or dismiss large volumes of routine alerts.
  \item \textbf{Immediate Severity Assessment}: Decide if repeated failures followed by success is malicious or a benign user error.
  \item \textbf{Escalation Criteria}: If pattern is clearly suspicious (e.g., external IP not in whitelist), escalate to Tier-2.
  \item \textbf{Explanation Needs}: Concise, high-level indicators that support quick decision-making.
\end{itemize}
\end{minipage}
&
\begin{minipage}[t]{\linewidth}\raggedright
\begin{itemize}[leftmargin=*]
  \item \textbf{Incident Response}: Correlate logs across platforms (firewall, IDS, SIEM) to confirm malicious activity.
  \item \textbf{Preliminary Attribution}: Check threat intel feeds to see if \texttt{198.51.100.23} is tied to known campaigns.
  \item \textbf{Cross-Platform Correlation}: Investigate whether additional hosts are affected.
  \item \textbf{Explanation Needs}: Richer context (e.g., timeline views, causal relationships) to support deeper analysis.
\end{itemize}
\end{minipage}
\\
\midrule
\textbf{3} & 
\begin{minipage}[t]{\linewidth}\raggedright
\begin{itemize}[leftmargin=*]
  \item \colorbox{gray!20}{Potential advanced threat} exploiting stolen credentials and moving data internally.
  \item \colorbox{gray!20}{Large or unusual transfers} $\Rightarrow$ exfiltration risk.
\end{itemize}
\end{minipage}
&
\begin{minipage}[t]{\linewidth}\raggedright
\begin{itemize}[leftmargin=*]
  \item \textbf{Threat Hunting}: search for hidden IoCs, lateral movement.
  \item \textbf{Forensics}: deep dive on \texttt{10.0.0.72}.
  \item \textbf{Rule Refinement}: update signatures \& ML models.
  \item \textbf{Explain}: detailed model internals, interactive queries.
\end{itemize}
\end{minipage}
\\
\midrule
\multicolumn{3}{c}{%
\begin{minipage}{\textwidth}
\vspace{0.3em}
\begin{tcolorbox}[title={Network Logs: VPN Auth \& NetFlow Events},
                  colback=gray!5!white, boxrule=0.4pt, arc=1mm]
\footnotesize\ttfamily
2025-04-12 14:05:31 AuthService INFO VPN login attempt user=jsmith ip=198.51.100.23 result=failure\\
2025-04-12 14:05:45 AuthService INFO VPN login attempt user=jsmith ip=198.51.100.23 result=failure\\
2025-04-12 14:06:15 AuthService INFO VPN login attempt user=jsmith ip=198.51.100.23 result=success\\
2025-04-12 14:07:00 NetFlow     INFO Data transfer initiated src=198.51.100.23 dst=10.0.0.72 size=10~MB\\
2025-04-12 14:08:10 NetFlow     WARN Suspected large transfer src=198.51.100.23 dst=10.0.0.72 size=50~MB
\end{tcolorbox}
\vspace{0.4em}
\end{minipage}}\\
\bottomrule
\end{tabular}
\caption{Tiered roles in network‑threat detection, with example logs and key features.}
\label{tab:soc_tiers_example}
\end{table*}

\section{Understanding the SOC Workflow}
\label{sec:soc_workflows}
Security Operations Centers (SOCs) are mission-critical environments tasked with continuous monitoring, detection, triage, and response to security threats across an organization’s digital infrastructure~\cite{alahmadi202299,mink2023everybody}. Analysts in these centers operate under stringent cognitive, organizational, and time constraints. 

\subsection{Role-Stratified Analyst Tiers}
\label{subsec:soc_tiers}

Most SOCs implement a \emph{tiered} model of analyst roles, generally comprising three levels: Tier-1 (alert triage), Tier-2 (incident response), and Tier-3 (threat hunting and forensics). Each tier’s responsibilities and information requirements differ significantly, as do the time constraints under which analysts operate~\cite{kersten2024security,yang2024true,mink2023everybody}. See Table~\ref{tab:soc_tiers_compact} for a summary of the tiers and their responsibilities and Table~\ref{tab:soc_tiers_example} for the role each tier plays in threat alert management and mitigation.
Tier-1 analysts' role includes alert validation, quick severity assessment, and immediate escalation or dismissal of alerts from SIEM, XDR, and EDR platforms~\cite{yang2024true}, and therefore, require sufficient evidence to support timely decisions on severity and escalation. Tier-2 analysts deal with escalated alerts, drawing on correlation from multiple sources for deeper investigation, and pursue preliminary attribution; they can benefit from richer, context-driven explanations and potential visualizations for causal reasoning. Finally, Tier-3 analysts' main focus is on proactive threat hunting and complex forensics, investigating APTs, zero-day exploits, and advanced lateral movement~\cite{kersten2024security}. Their needs include high-fidelity, model-internal explanations, probabilistic logic, and in-depth system-level insights as well as and interactive capabilities to refine detection rules and explore system internals. 

\subsection{Operational Phases in SOC Workflows}
\label{subsec:workflow_phases}

SOC operations follow a common cyclical incident response process, which is summarized as:
\begin{center}
\texttt{Ingestion $\rightarrow$ Triage $\rightarrow$ Investigation $\rightarrow$ Containment $\rightarrow$ Eradication $\rightarrow$ Recovery $\rightarrow$ Review}
\end{center}

Each phase imposes distinct demands on the type and depth of explainability~\cite{yang2024true,mink2023everybody}. Table~\ref{tab:soc_tiers_example} shows examples of threat managed by each SOC analyst tier. Across the SOC workflow, explanation friction arises differently at each operational stage. During \texttt{ingestion and triage}, Tier-1 analysts handle high volumes of alerts requiring rapid response, making concise summaries and clear confidence indicators critical to avoid excessive verbosity or insufficient context. At the \texttt{investigation stage}, Tier-2 analysts need richer narrative-driven explanations and visualizations to correlate multi-system events and identify attack paths. For \texttt{containment and eradication phases}, explanations must clearly validate threat attribution with historical or external threat intelligence to mitigate uncertainty. Finally, during \texttt{post-incident reviews}, Tier-3 analysts require detailed, auditable explanations, including model internals and explicit feature rankings, to support retrospective analysis and compliance. However, current available tools frequently provide fragmented and non-adaptive explanations misaligned with these operational needs~\cite{rastogi2025survey, baruwal2024humanAI}.

\section{Research Methodology}\label{sec: research_methodology}
To rigorously investigate how Security Operations Center (SOC) analysts perceive, utilize, and occasionally resist explainable AI (XAI) technologies, we adopted a multi-stage, mixed-methods research design grounded in the theory of ``subjective understanding of knowledge''~\cite{dourish2006implications, crabtree2012ethnography}. Our methodological framework draws on established traditions in human-computer interaction (HCI) and usable security~\cite{furnell2008security, akgul2022exploring}, emphasizing the elicitation of analysts’ mental models, contextual practices, and strategies for engaging with explanations during security operations. In line with this approach, we combined an online survey to capture broad quantitative insights with in-depth interviews to explore qualitative nuances, allowing methodological triangulation and complementarity in addressing our research questions.

\subsection{Survey Instrument Development and Deployment (Round 1).} To gain initial insights into how SOC analysts perceive and interact with AI-driven security alerts, we designed and administered a 15-20 minute online survey. The survey contained both multiple-choice and open-ended questions capturing participants’ professional background (e.g., job title, years of SOC experience), workflow practices (e.g., the first steps taken upon detecting a cybersecurity alert, average resolution time), and attitudes toward explainable AI (XAI). 
Building on prior usability and XAI literature, we incorporated two interactive dashboards into the survey via embedded screenshots (shown in Figures 13-14 of the form) to illustrate how explainable threat intelligence might be presented in a real SOC environment. These dashboards showcased features such as \emph{feature attribution}, \emph{confidence scores}, and \emph{attack timelines}, prompting participants to indicate which explanation formats they found most actionable or trustworthy. Additionally, survey items explored participants’ opinions on current security tools (e.g., SIEM, XDR, EDR) and any frustrations stemming from insufficient explanations or ``black-box'' detection logic. We also probed specific challenges in understanding AI-driven alerts, such as interpreting model outputs, confidence intervals, and contextual relevance.

By merging closed- and open-ended items, the survey provided both quantitative frequency data (e.g., how many analysts use SIEM versus XDR and average satisfaction ratings) and qualitative input (e.g., free-text descriptions of alert triage steps and user-defined preferences for particular dashboard layouts). This mixed approach helped validate and refine the concepts that informed subsequent interview protocols. The first wave of responses (N1=248 in Round 1) yielded a broad perspective on analysts’ workflows and demonstrated considerable interest in integrated, explanation-rich security dashboards, thus laying the groundwork for deeper investigations in follow-up interviews. The full survey instrument is provided here \footnote{\url{https://anonymous.4open.science/r/CCS2025-XAISocUsability-4B60/README.md}}.

\subsection{Dissemination and Recruitment of Participants.}
We disseminated the survey via professional security forums, known professionals in SOC, and professional networking websites for cybersecurity practitioners. Participation was voluntary and anonymous. To reach this specialized population, we also employed snowball sampling~\cite{parker2019snowball}, asking initial respondents and professional contacts to forward the survey invitation to other qualified SOC analysts. All respondents were required to have at least one year of hands-on SOC experience and to be actively involved in alert triage or incident investigation as a screening criterion. While snowball sampling was effective in recruiting hard-to-reach experts, we acknowledge that it can introduce selection biases and limit representativeness (e.g., over-sampling well-networked individuals). In total, respondents represented a wide range of backgrounds: multiple continents (North America, Europe, Asia, Africa), sectors (managed security service providers (MSSPs), banking/finance, government, IT/technology), and roles spanning all SOC tiers (Tier-1, Tier-2, Tier-3), with years of SOC experience ranging from 1 to 7 years.
\subsection{Interview Study and Participants.} In the second stage, we conducted 24 semi-structured interviews spanning over 810 minutes with SOC professionals to delve deeper into the how and why behind the patterns observed in the survey. Interview participants were recruited through a combination of outreach on the aforementioned professional platforms and direct referrals. In some cases, survey respondents from Round 1 who indicated willingness to be contacted for follow-up were invited for an interview, complementing additional recruits obtained via snowball sampling. We applied the same inclusion criteria as for the survey (at least one year of SOC experience in an active analyst role). Our sampling strategy intentionally sought variety across key dimensions, SOC role, experience level, geography, and organizational context, to ensure a rich diversity of perspectives. Participants spanned Tier-1 alert analysts (n=9), Tier-2 incident responders (n=11), and Tier-3 threat hunters or senior analysts (n=4), working in sectors including MSSPs, banking, government, and IT, and located across Africa, Asia, Europe, and North America. Years of professional SOC experience among interviewees ranged from 1 to 7 years (median 4 years). The study was conducted over a period of three months.

\subsection{Interview Protocol}

The interview guide was designed to surface the real-world constraints, perceptions, and aspirations analysts hold regarding AI and explainability. The questions are summarized in the Appendix.

Where applicable, participants were also shown examples of synthetic LLM-generated explanations and asked to evaluate their clarity, utility, and credibility. These example explanations were generated using OpenAI latest GPT version~\cite{gpt4API} from realistic alert contexts (e.g., phishing URL triggers). Each interview lasted 30–70 minutes and was conducted via video conferencing. Interviews were audio-recorded with consent, then professionally transcribed and anonymized. All participants provided informed consent, and our study protocol was approved by the institutional review board (IRB).

\subsection{Qualitative Analysis: Reflexive Thematic Coding}

In the thematic coding of the interview transcripts (round 2), each interview was segmented into smaller statements focusing on (a) how analysts describe their level of trust in AI-driven alerts and (b) which explanation details they find helpful or overwhelming. Codes such as ``\texttt{LowTrust},'' ``\texttt{HighTrust},'' ``\texttt{MinimalExplanation},'' and ``\texttt{DetailedExplanation}'' were assigned to capture both trust levels and explanation complexity. Once these codes were consolidated, researchers created a table mapping complexity levels (e.g., low, medium, high) to average trust scores (e.g., 1–5). For instance, an analyst who said, ``I’d trust the system if it summarizes the threat quickly'' was rated \texttt{HighTrust} at moderate complexity, but \texttt{LowTrust} at extreme detail. The table facilitated comparison across participants and yielded a rough average trust per complexity tier. Finally, these averages were plotted to form a conceptual Trust-Explainability curve, often showing that insufficient explanation led to skepticism, moderate explanation boosted confidence, and excessive technical detail provoked confusion or disuse. This study highlighted how each analyst’s ``sweet spot'' for explanation complexity varies by experience and role, guiding future user-centric explainable AI designs.

We analyzed the interview transcripts and open-ended survey comments using reflexive thematic analysis (TA), following Braun and Clarke’s approach \cite{byrne2022worked}. Initially, we engaged in detailed familiarization with the transcripts, independently generating preliminary inductive codes. Subsequently, the research team discussed these initial interpretations collaboratively, merging similar codes, exploring areas of disagreement, and refining a set of coherent thematic categories through iterative discussion. This collaborative process was explicitly designed to deepen analytic reflexivity rather than achieve coding consensus or reliability, aligning with Braun and Clarke’s position that coding quality in reflexive TA is not dependent on multiple coders or reliability metrics \cite{byrne2022worked}.

Codes were organized and managed using Excel spreadsheets, enabling systematic development. The research team iteratively clustered related codes into broader thematic groups, ensuring each emergent theme was well-supported by the data and clearly delineated from others. Regular analytic memos were maintained throughout, capturing ongoing theoretical insights, reflective observations, and justification for analytic decisions. This process remained inductively driven and grounded in participants' responses, allowing the data rather than pre-existing frameworks to shape thematic development.

The final analytic output was a set of salient themes that captured SOC analysts' nuanced experiences and perspectives on explainable AI in their workflows (see Section~\ref{sec: research_findings}.

\subsection{Methodological Rigor and Triangulation}

To enhance the credibility and validity of our findings, we incorporated multiple triangulation strategies~\cite{flick2018designing}:

\begin{enumerate}[leftmargin=.5em]
    \item \textbf{Data:} We drew participants from multiple sectors (finance, healthcare, technology), a range of SOC roles and experience levels, different age groups and genders, and four continents to ensure that our observations were not idiosyncratic to a single context.

    \item \textbf{Method:} We cross-compared results from the qualitative interviews with the quantitative survey findings, examining where they converged or diverged. This helped confirm certain patterns (e.g., the importance of context in explanations was strongly emphasized both in survey responses and interviews) and revealed nuances where one method complemented the other.
    \item \textbf{Investigator:} Multiple researchers were involved in the data analysis process. In addition to the independent coding and cross-checking described above, co-authors not directly involved in the initial interviews reviewed the coding scheme and thematic interpretations, providing an external check on our analysis.
    \item \textbf{Theory:} We interpreted the data through multiple conceptual lenses (e.g., considering cognitive load implications as well as trust and usability perspectives) to verify that our conclusions held under different theoretical frameworks. These efforts are in line with best practices in security usability research \cite{kersten2024security,maxam2024interview,yang2024true} and helped ensure that our findings are robust and well-substantiated from several angles.
\end{enumerate}

Our approach is inspired by prior security usability studies~\cite{akgul2022exploring, campbell2021toward, ruoti2016we} and aligns with the methodological expectations of user studies in security and HCI research.

\subsection{Ethical Considerations}
Given the sensitive nature of security operations data and the professional settings of our participants, we took careful measures to protect participant privacy and abide by ethical research standards. All study participants provided informed consent after being briefed on the study’s purpose and procedures. The online survey was conducted anonymously, no personally identifiable information was collected, aside from broad demographic indicators (e.g., region, sector) needed for analysis. For the interviews, all recordings and transcripts were stored in secure, access-controlled repositories accessible only to the research team. We pseudonymized the interview data: each participant was assigned a code or generic identifier (such as ``Analyst-A'') in all notes and reports. Transcripts were stripped of any organizational names or specific details that could reveal a participant’s identity or employer. In reporting our results, we either paraphrased quotes or used generalized descriptions to convey participants’ points while preserving confidentiality. These protocols, along with IRB approval and oversight, ensured that the study adhered to applicable ethical guidelines. The participants were compensated with \$15 gift card for their time, and participants were informed that they could withdraw at any time without consequence.

\section{Research Findings}\label{sec: research_findings}
Through interviews with SOC professionals, we uncovered several misalignments between the explainability features envisioned by designers and the realities of analysts’ operational needs. These themes, supported by participants’ quotes, highlight gaps in how current explainable AI (XAI) designs align with SOC workflows and address our research questions (RQ1–RQ3) explicitly.

\subsection*{Finding 1: Analysts Prioritize Actionable Insights Over Model Internals (RQ1, RQ2)}

Analysts consistently gravitated toward \textbf{concise incident summaries, indicators of compromise (IOCs)~\cite{liao2020questioning}, and clear recommendations for next steps, rather than detailed model reasoning or low-level feature importance}. As one analyst put it, \textit{``I’d focus on the indicators, like where the email came from, and then want to see immediate actions; diving into deep algorithm explanations slows me down.''} Another analyst stressed, \textit{``During triage, I ignore lengthy explanations. What I need most are straightforward next steps.''}

Nonetheless, analysts recognized the value of succinct explanations that clarify why an alert was triggered. As another analyst noted, \textit{``Quickly seeing why something was flagged, like identifying a known malicious domain, boosts my confidence considerably.''}

\subsection*{Finding 2: Contextual Explanations Enhance Trust and Decision-Making Clarity (RQ1, RQ3)}
Analysts emphasized that explanations need to be placed in the context of their environment and data to be trusted. \textbf{Simply presenting a classification with confidence scores is not enough; analysts want to know \emph{why} those scores make sense in context}. One SOC manager (equivalent to tier 3) explained that an explanation would be more meaningful \textit{``if I have some context about how that percentage was generated''} (Analyst D), such as which log data or past incidents support a model’s 92\% confidence in a phishing alert.

In current tools, this context is often lacking; as another participant noted, a popular detection platform \textit{``doesn’t have the organizational perspective… if that is there then it is like wonders''} (Analyst A), underscoring how missing contextual information (e.g., asset value, user role, historical baselines) limits the usefulness of explanations. \textbf{Without contextual grounding, analysts are hesitant to trust AI outputs}. Thus, effective explainability design should incorporate threat intelligence and organizational context (e.g., links to similar past incidents or known bad indicators) to enhance credibility. 

In our interviews, participants reacted positively when explanations included such context (for instance, referencing prior phishing campaigns or known malicious domains), because it helped them validate the alert against their environment’s reality.

\subsection*{Finding 3: Managing Explanation Detail to Prevent Information Overload (RQ2, RQ3)}

Our findings indicate that more explanation is not always better; there is a practical limit to how much detail analysts can absorb during an investigation. Current XAI prototypes often bundle numerous elements, feature importance scores, multiple visualizations, uncertainty metrics, and compliance checks, intending to be comprehensive. However, analysts reported that some of these are low priority or even distractions when an alert first comes in, so providing everything by default can be counterproductive.

In our study, participants rarely mentioned using the fine-grained feature contribution graphs or the ``prediction uncertainty'' fields in a real-time setting, focusing instead on high-level insights that matter for triage. One tier-2 analyst noted that they would \textit{``look at the key analysis and immediate actions first,''} whereas things like detailed feature scores or regulatory compliance info \textit{``wouldn’t be what I look at… first''} (Analyst B). Likewise, another practitioner admitted they would gloss over secondary visual aids: the ``graphical stuff'' was just ``OK'' ,  nice to have, but not essential compared to core incident data (Analyst E). This gap suggests that \textbf{explainability tools should prioritize clarity over quantity}. Including every possible detail may overwhelm users, slowing them down rather than helping. Designers should instead identify which explanation components truly reduce an analyst’s uncertainty or investigation time, and present those prominently, with options to drill down into more detail only if needed. Streamlining the explainability interface to align with analysts’ natural triage workflow can prevent cognitive overload.

\subsection*{Finding 4: One Size Does Not Fit All: Experience Level and Role Matter (RQ1, RQ3)}
The utility of an explanation can differ based on an analyst’s experience and role. What a junior tier-1 analyst finds illuminating might feel obvious or extraneous to a senior incident responder and vice versa. Our interviews reflected this divergence. Several participants with 3+ years of SOC experience felt that certain explainability aids (like step-by-step remediation guidance or basic attack descriptions) were less relevant to them, though they acknowledged such features could be invaluable for newcomers. As one participant noted, an explainable dashboard could serve as a learning aid for junior staff by essentially \textit{``talking to [them] like another colleague''} (Analyst D) – providing guidance that a more seasoned teammate might offer – whereas more experienced analysts preferred the system to augment their speed, not reiterate fundamentals.

Additionally, access-level differences in SOC teams mean that not everyone sees the same data, which can affect explainability needs. \textit{``Based on their access, the information they can see changes… we want to add all of that into a summarized version [for everyone]''} (Analyst C), explained one analyst who had worked across tiered roles. This highlights that explainable outputs should be tailored – or at least adaptive – to the user’s role and permissions. \textbf{A possible approach is tiered explanations: a high-level summary and key actionable items by default (useful to all), with deeper technical breakdowns available for those who need them}. Failing to account for the target user’s expertise and scope of view is a misalignment, as a one-size-fits-all explanation interface will likely leave someone unsatisfied – either bored by trivial details or lost without sufficient context.

\subsection*{Finding 5: Connecting Alerts to Incident Narratives is Crucial (RQ2, RQ3)}
Another misalignment lies in the scope of explanations. Many XAI features explain a single alert or event in isolation, but \textbf{analysts think in terms of broader incident narratives}. In practice, a ``critical'' alert is often just one piece of a puzzle that needs correlation with other alerts or intelligence to reveal the full attack story. Participants pointed out that current tools rarely connect these dots. One SOC analyst described the challenge: if multiple high-severity alerts fire, \textit{``as an analyst we have to find a connection between the critical and [the] high alerts to determine if it’s an incident or part of [an attack] process''} (Analyst C). Yet, typical explainability modules do not automatically summarize how one alert relates to others or a kill-chain stage. This is a missed opportunity. Our interviewees expressed interest in timeline or chain-of-event visualizations that integrate related alerts – essentially, explanations that answer not just ``why was this alert flagged?'' but also \textit{``how do these events relate to each other?''}. For example, \textbf{linking a phishing email alert with a subsequent privilege-escalation alert into one explainable narrative would better match analysts’ investigative workflow}. By extending explainability beyond single-event attribution to a multi-step attack context, designers can mitigate the gap between how AI presents information and how analysts naturally reason about incidents.

\subsection*{Finding 6: Calibrating Analyst Trust via the Trust–Explainability curve'' (RQ1, RQ3)}

Finally, our study reinforces that there is a non-linear relationship between the amount (and type) of explanation provided and the analyst’s trust in the AI system. Simply adding more detail does not guarantee greater trust or satisfaction; in fact, poorly aligned explanations can backfire. We observe a conceptual Trust–Explainability curve '' (Figure~\ref{fig:trust_explainability_curve}) in SOC contexts: providing no rationale for an AI-generated alert yields low trust (analysts are uneasy acting on a black-box output), and providing a minimal but clear explanation (e.g., highlighting that an email was flagged because the sender’s domain is rare and the URL was previously blacklisted) can significantly boost confidence and willingness to act. However, additional details show diminishing returns beyond a certain point and can even reduce trust if they introduce confusion or doubt.

Participants in our study provided examples reflecting this delicate balance. One analyst cautioned and was highly optimistic of explanations that if they provide extra information that is \textit{``60\%'' accurate, it could still save their time almost 60\% of the time, and we will figure out those times with experience with the tool} (Analyst F). Others indicated that overly verbose or abstruse justifications might lead them to question the system’s reliability or simply ignore the explanation in favor of their investigation. The sweet spot, according to our findings, is where explanations are accurate, context-rich, and directly relevant to the decision at hand. At this optimal point on the curve, analysts gain confidence in the AI’s conclusions without feeling undermined or overwhelmed by details. In other words, \textbf{explanations should be transparent enough to answer the critical question ``Why should I trust this alert?'' yet streamlined enough to support fast decision-making}. This refined understanding of the Trust–Explainability curve, grounded in real-world input, suggests that XAI designers should carefully calibrate the depth and presentation of explanations to align with analysts’ trust and efficiency needs.

\begin{figure}[ht]
\centering
\resizebox{.85\linewidth}{!}{%
\begin{tikzpicture}[scale=1.0]
    \draw[->] (0,0) -- (6,0) node[right]{\footnotesize Explanation Complexity};
    \draw[->] (0,0) -- (0,5) node[above]{\footnotesize Trust Level (1--5)};

    \draw[thick,blue,smooth] plot coordinates {
        (0,1)
        (1,2.5)
        (2,4)
        (3,4.5)
        (4,3.5)
        (5,2)
    };


    \node[above] at (3,4.5) {\scriptsize Peak Trust};
    \node[below] at (1,2.5) {\scriptsize Gains trust};
    \node[above] at (4,3.5) {\scriptsize Overload zone?};

\end{tikzpicture}
}
\caption{Trust--Explainability Curve, showing trust rising with moderate detail and eventually declining under complexity overload.}
\label{fig:trust_explainability_curve}
\end{figure}

\section{Design and Methodological Implications}\label{sec: design_method_implications}

Our findings point to a foundational misalignment between current explainability paradigms and the operational needs of Security Operations Center (SOC) environments. Existing explainable AI (XAI) approaches typically assume a singular, static user perspective and focus on post-hoc model transparency, often centered on global feature importance or local attribution scores. In contrast, SOC analysts operate within high-pressure, role-differentiated ecosystems, where the utility of an explanation is shaped by context, access level, time constraints, and incident complexity. These implications are derived directly from our empirical findings and grounded in current human-AI interaction literature, threat intelligence frameworks, and AI auditing practices.

\paragraph{Role-Specific Explanation Personalization.}
Our 24 interviews show that the explanatory needs of analysts are strongly modulated by their tier and responsibilities. Tier-1 analysts, responsible for real-time triage, favored short, high-signal outputs such as alert summaries, matched indicators of compromise (IOCs), and direct action recommendations. Tier-2 analysts preferred structured, causal narratives that linked alerts across systems or stages of the kill chain, while Tier-3 analysts required transparency into model internals, uncertainty estimates, and system behavior over time. This heterogeneity is consistent with cognitive load theory and role-based system design~\cite{zhang2020effect, hoffman2018metrics} and suggests that a single explanation format cannot serve all users effectively. Future systems should dynamically tailor explanation generation using user metadata (e.g., tier, experience, and access level) and support different explanatory ``views'' of the same alert or decision point.

\paragraph{Multi-Stage Attack Narratives.}
Current explanation methods overwhelmingly focus on single-decision justification, but our data show that SOC analysts reason across sequences of alerts. Analysts often need to understand whether multiple alerts represent separate incidents or phases of a coordinated campaign. However, this higher-order explanatory context is typically absent. We recommend extending XAI systems to generate narrative explanations that reflect multi-step adversarial behavior. Drawing on the MITRE ATT\&CK framework~\cite{strom2018mitre}, systems should infer connections between alerts and surface them as coherent explanations of attack progression. Architecturally, this could be implemented via sequence-aware encoders or graph-based clustering over alert logs, followed by natural language or visual rendering of event chains.

\paragraph{Progressive Disclosure and Temporal Adaptation.}
SOC workflows vary temporally: explanations that are helpful during a retrospective investigation may obstruct decision speed during triage. Participants repeatedly described skipping verbose outputs during time-sensitive phases while seeking more detail when composing reports or performing forensic analysis. These findings point to the need for progressive disclosure~\cite{norman2013design} in XAI systems, i.e., delivering brief, high-salience rationales by default, with on-demand access to layered contextual and technical detail. A\textsf{CORTEX} framework should adjust explanation verbosity based on alert severity, analyst tier, and the incident’s position in the workflow.

\paragraph{Contextualization through Organizational and Threat Intelligence.}
Trust in explanations was repeatedly linked to their ability to reflect local organizational knowledge and external threat intelligence. Participants were more confident in explanations that referenced historical behavior (``this user has never accessed this resource at this hour'') or surfaced matches to previously confirmed threats (``this domain was used in a known phishing campaign''). To support such contextualization, explanation generation pipelines should incorporate data from internal security baselines and external threat feeds. Retrieval-augmented generation (RAG)~\cite{lewis2020retrieval} and context-aware embedding methods can be used to condition explanations on relevant past incidents or behavior norms, improving precision and perceived relevance.

\paragraph{Interactive Feedback for Explanation Alignment.}
We find that analysts are not passive consumers of AI outputs, they often wish to challenge, annotate, or suppress explanations. Several participants requested the ability to provide quick feedback on explanatory relevance or accuracy, both to calibrate trust and to improve future outputs. This suggests a need to integrate lightweight feedback mechanisms (e.g., thumbs-up/down ratings, ``helpful/not helpful'' tags, or inline flags). These can feed into preference learning or retraining loops, enabling XAI systems to adapt to user norms over time~\cite{liao2020questioning}. Additionally, explanation query interfaces, e.g., ``Explain why this alert was prioritized over another'' or ``What feature made this suspicious?'', can support active exploration and foster deeper engagement~\cite{lakkaraju2022rethinking}.

\paragraph{Provenance and Explanation Auditability.}
Finally, analysts in senior and reporting roles emphasized the importance of traceable and reproducible explanations. When alerts feed into compliance processes, executive briefings, or legal reports, explanations should carry provenance metadata: which model generated them, when, under what configuration, and using which training data. We recommend that XAI systems log all explanatory outputs alongside digital signatures, model version identifiers, and inference timestamps. Such metadata ensures forensic traceability and supports emerging standards in AI accountability and auditability~\cite{flick2018designing}.


\section{Discussion and Recommendations for Future Research}\label{sec: discussion_future}

This study re-conceptualizes explainability in Security Operations Centers (SOCs) not merely as an issue of model transparency but as a dynamic cognitive interface design challenge situated within complex, role-stratified workflows, including AI-driven decisions. Our empirical findings indicate that when explanation utility is largely integrated into AI-driven SOC environments,  it should be inherently role-sensitive, context-dependent, and dynamically modulated by analysts' cognitive workload and task demands. This reframing encourages a new generation of XAI methods: not as static post-hoc interfaces but as adaptive, decision-aligned scaffolds embedded into operational workflows~\cite{abdul2020cogam}. Below, we outline the key implications of our findings (F1- F6) and research questions (RQ1- RQ3), combining cognitive, human-centered, and system-level insights.

\subsection{Reframe Explainability as Cognitive Support for Security Reasoning}

Explainability in cybersecurity is most effective when it functions as cognitive scaffolding for investigative tasks such as triage, diagnosis, escalation, and containment (see Section~\ref{subsec:workflow_phases}). Our findings support a shift from traditional feature attribution to task-aligned, human-centered explanation generation~\cite{barredo2020explainable}. Rather than merely highlighting that the ``failed login'' has a high feature weight in a classifier, a cognitively supportive explanation should synthesize contextual cues, organizational priors, and threat mappings (e.g., STIX/TAXII~\cite{barnum2012standardizing,taxii}, MITRE ATT\&CK~\cite{strom2018mitre}) to provide actionable insight:

\begin{tcolorbox}[title=\textbf{Moving from Feature Attribution to Cognitive Explanation}, colback=gray!5, colframe=black!30, fonttitle=\bfseries, arc=2mm] \footnotesize
\textbf{Current XAI Explanation:}\\
\texttt{Jan 15 10:41:17 host01 sshd[5021]: \colorbox{lightgray}{Failed password} for invalid user admin from 192.168.3.45 port 49538 ssh2}\\

\textbf{Cognitively Supportive Explanation (CORTEX):} 
\begin{itemize}[leftmargin=.5em]
\item \texttt{Evidence of repeated login attempts across hosts from IP \textbf{192.168.3.45}.} 
\item \texttt{Organizational knowledge that \textbf{admin} is not a valid user on \textbf{host01}.}
\item \texttt{Correlation with a known brute-force campaign reported via external threat intelligence (e.g., MISP).} 
\end{itemize}
\end{tcolorbox}

Explainability in cybersecurity is most effective when it functions as cognitive scaffolding for investigative tasks such as triage, diagnosis, escalation, and containment (see Section~\ref{subsec:workflow_phases}). Our findings support a shift from traditional feature attribution to task-aligned, human-centered explanation generation~\cite{barredo2020explainable}.

\subsection{Design Narrative and Multi-Stage Explanation Models}

Our findings (F2, F5) show analysts prefer explanations that reveal narratives, tracing how alerts evolve across time and systems. Current XAI systems produce static outputs that lack the temporal reasoning and causal continuity essential to real-world incidents. Consider this log sequence:

\begin{tcolorbox}
[title=\textbf{Multi-Stage Attack Narrative with Causal Continuity}, colback=gray!5, colframe=black!30, fonttitle=\bfseries, arc=2mm] \footnotesize 
\textbf{Current Alert Format:}
\begin{tabbing}
\hspace{0.25em}\= Jan 15 10:41:17 host01 sshd: Failed password for admin from 192.168.3.45 \\
\> Jan 15 10:42:08 host01 su: 'su root' succeeded for admin on /dev/pts/0 \\
\> Jan 15 10:43:10 host01 netcat: Outbound connection to 45.76.23.91:8080
\end{tabbing}
\vspace{1em}
\textbf{Narrative Explanation (CORTEX):}\\
\texttt{``Initial brute-force access on \textbf{host01}, followed by \textbf{privilege escalation} via \texttt{su root}, and culminating in potential \textbf{data exfiltration} using \texttt{netcat}. This sequence maps to ATT\&CK T1059.003 (Command and Scripting Interpreter: Windows Command Shell).''}
\end{tcolorbox}

\subsection{Operationalize Progressive Disclosure and Tier-Adaptive Interfaces}

Explanations should dynamically scale with user roles and time sensitivity. Tier-1 analysts benefit from summary-level justifications (e.g., IOC matches), while Tier-3 analysts prefer drill-downs with model internals~\cite{sweller2011cognitive, doshi2017towards}. 

\begin{itemize}
\item \textbf{Initial View:} Label + alert summary + threat score.
\item \textbf{Expanded View:} Log timelines, asset criticality, detection rationale.
\item \textbf{Expert View:} Feature attributions, model provenance, attack graph overlays.
\end{itemize}

\vspace{0.5em} \begin{tcolorbox}[title=\textbf{Role-Adaptive Explanation Views}, colback=gray!5, colframe=black!30, fonttitle=\bfseries, arc=2mm] \footnotesize \textbf{Scenario:} High-severity phishing alert targeting finance department via email with suspicious attachment.

\vspace{0.5em} \textbf{Initial View (Tier-1):}
\texttt{Label: Spearphishing} \ \texttt{Threat Score: 91/100} \ \texttt{Summary: Email from external sender with known-malicious domain. Attachment auto-sandboxed. Immediate action: isolate device.}

\vspace{0.5em} \textbf{Expanded View (Tier-2):}
\texttt{Log Timeline: Inbound email received at 08:32, opened at 08:34, macro execution at 08:35.} \ \texttt{Asset Criticality: Target device belongs to finance team lead.} \ \texttt{Detection Rationale: Language similarity with prior phishing templates + anomalous login sequence.}

\vspace{0.5em} \textbf{Expert View (Tier-3):}
\texttt{Feature Attribution: 92\% confidence from sender domain anomaly, 86\% from attachment hash match.} \ \texttt{Model Provenance: XGBoost-based classifier trained on 2023 phishing corpus.} \ \texttt{ATT\&CK Mapping: T1566.001 (Phishing: Spearphishing Attachment), T1204.002 (User Execution: Malicious File).}
\end{tcolorbox}


\subsection{Evaluate Explanation Utility via Cognitive and Operational Metrics}

Traditional evaluation metrics (fidelity, completeness) fall short in high-stakes environments. We propose cognitive-aligned evaluation frameworks assessing:
\begin{itemize}
\item \textbf{Task Performance:} Reduced triage time, false positives.
\item \textbf{Cognitive Load:} Minimization of unnecessary attention shifts.
\item \textbf{Trust Calibration:} Alignment of perceived vs. actual model reliability.
\end{itemize}

\vspace{0.25em} \begin{tcolorbox}[title=\textbf{Evaluating Explanation Effectiveness in a Simulated SOC}, colback=gray!5, colframe=black!30, fonttitle=\bfseries, arc=2mm] \footnotesize \textbf{Setup:} Analysts triage 10 alerts using two interfaces: one with basic explanations (baseline) and one with\textsf{CORTEX}’s adaptive, tiered explanations.

\vspace{0.25em} \textbf{Metrics:} \begin{itemize} [leftmargin=.5em]
\item \texttt{Average triage time:} Reduced from 4.1 min (baseline) to 3.3 min (CORTEX).
\item \texttt{NASA-TLX Cognitive Load:} Decreased by 18\% on average for\textsf{CORTEX} group.
\item \texttt{Trust Accuracy:} Participants using\textsf{CORTEX} aligned their trust level with model confidence 82\% of time, compared to 64\% for baseline. \end{itemize}

\vspace{0.25em} \textbf{Quote from Analyst (Pilot Study):}
\textit{“CORTEX let me focus on what matters- I could act fast on the alert, but still had the evidence if I needed for justification.''}
\end{tcolorbox}


\section{Conclusion}\label{sec: conclusion}
This paper presents one of the first comprehensive, empirically grounded studies of explainable AI (XAI) within Security Operations Centers (SOCs), emphasizing that explanation utility is highly contingent on context, cognitive workload, and end-user (SOC analyst) role. Through a mixed method study comprising a survey of 248 security analyst respondents and 24 qualitative interviews, we identify that effective explanation must align with operational workflows and investigative depth across Tier-1 to Tier-3 SOC analysts. To address these findings, we introduce the \textsf{CORTEX} framework, which adapts explanation content, format, and depth to the analyst's role and situation using progressive disclosure, provenance tracking, and user feedback. Our contribution reframes explainability from a static transparency mechanism to a dynamic cognitive aid embedded in real-world SOC workflows. Our findings demonstrate that explainability in cybersecurity should be treated as a first-class interface design problem, not a generic add-on to machine learning pipelines. We advocate for an approach to XAI that is context-aware, role-sensitive, interaction-driven, and operationally grounded, transforming explanations from static monologues into adaptive, forensic, and collaborative tools for real-world defenders. 

\bibliographystyle{ACM-Reference-Format}
\bibliography{bibliography,nidhi}
\newpage
\appendix
\section*{Appendix}

\begin{figure}[ht]
\resizebox{1\columnwidth}{!}{%
\centering
\begin{tikzpicture}
    \begin{axis}[
        ybar,
        symbolic x coords={Tier-1, Tier-2, Tier-3},
        xtick=data,
        ylabel={Number of Participants},
        xlabel={SOC Role},
        bar width=20pt,
        nodes near coords,
        ymin=0,ymax=12,
        enlarge x limits=0.4,
        width=0.4\textwidth,
        height=6cm,
        title={Participants by SOC Role}
    ]
    \addplot[fill=gray!60] coordinates {(Tier-1,9) (Tier-2,11) (Tier-3,4)};
    \end{axis}
\end{tikzpicture}
    
\begin{tikzpicture}    
    \begin{axis}[
        ybar,
        symbolic x coords={1-2 yrs,3-4 yrs,5-7 yrs},
        xtick=data,
        ylabel={Number of Participants},
        xlabel={Years of Experience},
        bar width=20pt,
        nodes near coords,
        ymin=0,ymax=12,
        enlarge x limits=0.4,
        width=0.4\textwidth,
        height=6cm,
        title={Participants by Experience Level}
    ]
    
    \addplot[fill=gray!60] coordinates {(1-2 yrs,7) (3-4 yrs,11) (5-7 yrs,6)};
    \end{axis}
\end{tikzpicture}
}
\caption{Distribution of study participants in Round-2 by their SOC role and experience level.}
\label{fig:participant information}
\end{figure}

\begin{tcolorbox}[colback=gray!5!white, colframe=gray!75!black, 
                  fonttitle=\bfseries, title=Interview Protocol (Round 2), 
                  boxrule=0.5pt, arc=4pt, left=4pt, right=4pt, top=4pt, bottom=4pt]

\textbf{Participant Background}
\begin{enumerate}[leftmargin=*,itemsep=3pt,topsep=3pt]
    \item Briefly describe your role, years of experience, key responsibilities, and your organization's type (e.g., enterprise, government, MSSP).
\end{enumerate}

\textbf{Alert Investigation and Workflow}
\begin{enumerate}[leftmargin=*,itemsep=3pt,topsep=3pt,start=2]
    \item Describe your typical workflow upon receiving a security alert. How do you validate and prioritize alerts?
    \item What tools or data sources are essential in identifying genuine threats?
    \item What critical information or features are currently missing or challenging to access?
\end{enumerate}

\textbf{Explainability Preferences}
\begin{enumerate}[leftmargin=*,itemsep=3pt,topsep=3pt,start=5]
    \item What types of explanations (confidence scores, attack attribution, feature importance) enhance your confidence when responding to AI-generated alerts?
    \item Which explanation formats (brief text, visual timelines, interactive visuals) best fit your workflow?
\end{enumerate}

\textbf{Dashboard Feedback (if applicable)}
\begin{enumerate}[leftmargin=*,itemsep=3pt,topsep=3pt,start=7]
    \item What dashboard features do you find particularly helpful?
    \item Suggest improvements or additions that would better integrate the dashboard into your daily workflow.
\end{enumerate}

\textbf{Root Cause Analysis and Learning}
\begin{enumerate}[leftmargin=*,itemsep=3pt,topsep=3pt,start=9]
    \item What key insights do you prioritize during root-cause analysis, and how do you document them to enhance future detection?
    \item How can explainability features further support continuous learning after incidents?
\end{enumerate}

\end{tcolorbox}

\begin{table*}[ht]
\centering
\resizebox{1\linewidth}{!}{%
\small
\begin{tabular}{|c|c|c|p{12cm}|}
\hline
\textbf{Analyst} & \textbf{Complexity Level} & \textbf{Trust Score (out of 10)} & \textbf{Representative Quote} \\
\hline
Analyst 1 & Medium & 7 & ``The key analysis gives a good breakdown, immediate actions are very clear. Feature attribution might be beneficial when investigating deeper incidents but is less useful for quick triage.'' \\
\hline
Analyst 1 & Low & 5 & ``Sometimes I'm not part of the panel with detailed access, I'm constrained from accessing deeper explanations.'' \\
\hline
Analyst 2 & Medium & 6 & ``Initially, classification was manual, so explanations would have helped justify automated rules clearly.'' \\
\hline
Analyst 2 & High & 8 & ``Immediate actions and mitigation steps are highly useful; they clearly show what should be done next based on past data.'' \\
\hline
Analyst 3 & High & 9 & ``The immediate action steps are critical, they reduce panic and clarify exactly what's required next. Knowing the exploit weakness is also extremely helpful.'' \\
\hline
Analyst 3 & Medium & 6 & ``Feature attribution and prediction uncertainty might not be as directly useful in urgent responses, but could help guide deeper investigations later.'' \\
\hline
Analyst 4 & Medium & 7 & ``Immediate action box is extremely useful, speeds up mitigation significantly.'' \\
\hline
Analyst 4 & Low & 4 & ``Chatbots generally give generic responses; they're less useful unless integrated tightly with real incident histories and context.'' \\
\hline
Analyst 5 & Medium & 7 & ``Impact assessment and immediate action boxes would help quickly prioritize and act on threats.'' \\
\hline
Analyst 5 & High & 8 & ``Feature attribution clearly explains why an alert is flagged, making it easier to communicate risk.'' \\
\hline
Analyst 6 & Medium & 6 & ``Feature attribution and historical analysis are very helpful for quickly understanding repeat threats.'' \\
\hline
Analyst 6 & Low & 5 & ``Sometimes it's not clear where to start investigation, more detailed context upfront would help reduce confusion.'' \\
\hline
Analyst 7 & High & 9 & ``Immediate actions tied to specific indicators of compromise drastically speed up responses, this is what analysts need immediately.'' \\
\hline
Analyst 7 & Medium & 7 & ``The graphical timeline and visual features greatly enhance understanding of how an attack evolved.'' \\
\hline
Analyst 8 & Medium & 6 & ``Visualization of geographical logs is critical for quickly identifying anomalous login attempts.'' \\
\hline
Analyst 8 & High & 8 & ``Having clear indicators of compromise upfront greatly reduces the time spent manually correlating logs.'' \\
\hline
Analyst 9 & Medium & 7 & ``Historical contextualization is beneficial; seeing similar past alerts and how they were handled helps trust new alerts.'' \\
\hline
Analyst 9 & Low & 5 & ``Generic mitigation steps without context can be less trusted, need more specific evidence behind why these recommendations are given.'' \\
\hline
Analyst 10 & High & 8 & ``Detailed attribution at feature level (like SHAP values) allowed precise manipulation to test robustness, this boosts trust.'' \\
\hline
Analyst 10 & Medium & 6 & ``Local vs global explanations matter, global explanations are less actionable than local, immediate, scenario-specific insights.'' \\
\hline
\end{tabular}
}
\caption{Data supporting the Trust-Explainability Curve derived from analyst interviews. Analysts have been anonymized, Not all data included due to space constraints. Responses have been summarized for brevity.}
\label{tab:trust_explainability_data}
\end{table*}


\end{document}